\documentclass[aps,eqsecnum]{revtex4}
\usepackage{latexsym}
\usepackage{graphicx}

\newcommand{\eq}{\!\!\!\! =\!\!\!\!}
\newcommand{\sigmalambda}{\bar\lambda}

\begin{document}
\title{\bf Unexorcized ghost in DGP brane world}
\author{
Keisuke Izumi$^{a}$\footnote{{\bf e-mail}: ksuke@tap.scphys.kyoto-u.ac.jp},
Kazuya Koyama$^{b}$\footnote{{\bf e-mail}: kazuya.koyama@port.ac.uk} and
Takahiro Tanaka$^a$\footnote{{\bf e-mail}: tama@tap.scphys.kyoto-u.ac.jp}}

\address{$^a$Department of Physics, Kyoto University, Kyoto 606-8502, Japan}
\address{$^b$Institute of Cosmology and Gravitation, University of
Portsmouth, Portsmouth PO1 2EG,UK}
\begin{abstract}
The brane world 
model proposed by 
Dvali-Gabadadze-Porrati realizes
self-acceleration of the 
universe. However, it is known that this cosmological solution
contains a spin-2 ghost. We study the possibility of avoiding the
appearance of the ghost by slightly modifying the model via the
introduction of a second brane. First, we consider a simple model
without stabilization of the brane separation. 
By changing the separation between the branes, we find that 
we can erase the spin-2 ghost. However, this can be done only at the
expense of the appearance of a spin-0 ghost instead.
We discuss why these two different types of ghosts are
correlated. Then, we examine a model with stabilization of the
brane separation. Even in this case, we find that the correlation
between spin-0 and spin-2 ghosts remains. As a result we find that 
we cannot avoid the appearance of a 
ghost by introducing a second brane in the model.
\end{abstract}
\maketitle

\section{Introduction}
\label{intro}
The observed 
present-day accelerated expansion of the universe~\cite{SN} is one
of the hottest topics in cosmology. There are two aspects in this
issue: The first is to explain the late-time accelerated expansion
of the universe, and the second is to explain why the accelerated
expansion began only recently. There have been many attempts to
modify standard cosmological models in order to address these
problems. As far as we know, none of these give a natural solution
to the second problem.  To give a solution to the second problem,
one may need to combine ideas which solve the first problem with the
anthropic principle~\cite{anth}.

In this paper, our interest is in the first problem. Most of ideas
that solve the first problem
are modifications of the scalar (spin-0) sector 
of the cosmological model, such as
introduction of a 
cosmological constant~\cite{const} or quintessence~\cite{quint}.
But there is another direction
which has not been explored much so far.
That is modifying the gravity theory in the spin-2 sector.

The simplest model with a modified spin-2 sector involves massive
gravity~\cite{FP}. Since models which have terms quadratic in
metric perturbations can be regarded as a massive gravity
theories, most modified gravity models fall into this category. If
we introduce the mass of the graviton to explain the accelerated
expansion of the universe, its value would be the same order as
the present value of the Hubble parameter, 
$m_g \simeq H$.
However, it is known that
a spin-2 graviton with mass in the range $0<m^2<2H^2$
in de Sitter background
has a ghost excitation in its helicity-0 component~\cite{Higuchi}.
The Hubble parameter is larger in an earlier epoch of the universe.
Hence it seems difficult to avoid the 
appearance of a ghost throughout the evolution of the universe in
simple models that attempt to explain the 
present-day accelerated expansion. Here we would like to raise a
simple question:  Can we build a model, incorporating modification
of the spin-2 sector, that explains the accelerated cosmic
acceleration and is ghost-free?

The DGP brane world model~\cite{DGP} (See Ref.~\cite{Lue} for a
recent review paper),
in which the 4D Einstein-Hilbert action 
is assumed to be induced on the brane,
is a mechanism to realize the late-time accelerated 
expansion of the Universe without introducing
additional matter~\cite{self}.
Such a solution is called the self-accelerating 
branch.  This is in contrast to the normal branch, where there is no
cosmic expansion in the limit of vanishing matter energy density.
There are many investigations of the DGP model, such as spherically
symmetric solutions~\cite{SSS,Lue2}, analyses of structure
formation~\cite{SF} and shock wave limits~\cite{SW}. In quantizing
the theory, we face the issue of strong coupling~\cite{strong,SG},
which is the problem that nonlinear effects of the scalar
gravitational interaction become important below some unexpectedly
large length scale.  Moreover, the existence of a ghost excitation
has been pointed out in the self-accelerating branch of the DGP
brane world model~\cite{SG,Koyama,Koyama2}. We here discuss this
ghost problem. By considering perturbations around this solution, a
Kaluza-Klein tower of massive gravitons in the four dimensional
effective theory is obtained. The lowest mass $m$ satisfies $m^2
=2H^2$~\cite{Koyama,Koyama2}. If the analogy to the massive gravity
theory holds,
this would mean that the self-accelerating branch 
of the DGP model does not have a ghost excitation. However, it a
detailed analysis has shown that there is still a ghost in this
model~\cite{Koyama2}.

In this paper, based on the DGP brane world model,
we attempt to build a ghost-free model which
simultaneously explains the accelerated expansion of the
universe due to effects in the 
spin-2 sector. The first idea is to increase the lowest graviton
mass by introducing a boundary brane.
We call this the two-brane model, which is successful in making
all graviton masses satisfy $2H^2<m^2$.  However, a spin-0
excitation, which originates from the brane bending degrees of
freedom, is transmuted into a ghost. We will find that a ghost
excitation cannot be eliminated in this two-branes model.

Usually spin-2 modes are decoupled from spin-0 modes at the level
of linear perturbations. Therefore it seems mysterious that
the disappearance of the 
spin-2 ghost coincides with the
appearance of a 
spin-0 ghost. The magic is in that a spin-2 mode can be obtained
from a spin-0 mode by applying a differential operator
if and only if the mass squared takes a 
specific critical value. At the critical mass, a spin-2 mode can
couple with a spin-0 mode. In addition, the critical mass
corresponds to a threshold for
the appearance of a ghost in the 
spin-2 sector. If the mass of a 
spin-2 mode is smaller than the critical mass, the 
spin-2 sector has a ghost excitation as previously mentioned.
Moreover, a unique spin-0 mode that appears in the
two-branes model has also the critical mass.
Hence the spin-2 mode can couple with the spin-0 mode when
it crosses the critical mass.
This fact partly explains
why the ghost can be transferred between
the spin-2 and spin-0 sectors. 

Then, we consider a model with stabilized brane separation
by introducing a bulk scalar field into the 
two-branes model~\cite{GW,TM}.
Since the mass spectrum of spin-0 modes 
changes, the 
spin-0 mode at the critical mass does not survive in this
stabilized two-branes model. Then, it is expected that the ghost
will not be transferred
between the spin-2 and spin-0 sectors. 
However, by studying this stabilized two-branes model in detail,
we will find that this naive expectation is wrong. 

This paper is organized as follows.
In \S\ref{twobrane} we will discuss the two-branes model.
In \S\ref{Stabilization} extension to the
stabilized two-branes model is performed.
We will summarize our results in \S\ref{Summary}.

\section{Two-branes model}
\label{twobrane}

As mentioned in Introduction, it is known that in de Sitter
spacetime a massive spin-2 field whose mass squared is less than
$2H^2$ has a ghost \cite{Higuchi}, where $H$ is the
Hubble parameter. 
The self-accelerating branch 
of the 
DGP brane world model \cite{DGP,self} has a spin-2 excitation whose mass squared
is $2H^2$. In the usual massive graviton model with Fierz-Pauli mass term,
this is a special case where a helicity-0 excitation disappears \cite{Higuchi, Deser}.
However, a detailed analysis
reveals that there is a ghost mode in the self-accelerating 
branch of DGP model due to the existence of a spin-0
excitation~\cite{Koyama2}. Since the ghost arises due to a subtle
mechanism in this model, it would be a natural question whether
one can eliminate the ghost by a slight modification of this model
or not.

In this section we examine the two-branes model with a boundary
brane in the bulk as one of the simplest extensions of original
DGP scenario.  Our initial expectation is as follows: The new
boundary will shift the mass of spin-2 modes. Then we will be able
to construct models with no spin-2 excitations with masses in the
range that implies the existence of a ghost.

However, we will find later that spin-0 excitations, which
correspond to the brane bending degrees of freedom, carry a ghost
mode instead. It will turn out to be impossible to eliminate the
ghost by this simple two-branes extension. Below, we discuss how
the ghost is transferred
between the spin-2 and spin-0 sectors~\cite{Kaloper}. 

\subsection{Background}

As mentioned above, in order to obtain a model that does not have
spin-2 ghost modes, we introduce another brane into the DGP model.
The action is
\begin{eqnarray}
S = { 1 \over 2 \kappa ^2} \int d^5x \sqrt{-g} R
     + \sum_{\sigma =\pm}  \int d^4x  \sqrt{-g^{(4)}_\sigma}
     \left( { 1 \over 2 \kappa_4^2} R^{(4)}_\sigma + {1 \over \kappa^2
      } K_\sigma - \tau_\sigma + L_{m\sigma} \right), 
\end{eqnarray}
where $R$, $R^{(4)}_\pm$, $g^{(4)}_\pm$, $K_\pm$, $\tau_\pm$ and
$L_{m\pm}$ are the five dimensional Ricci scalar, four dimensional
Ricci scalar, trace of the four dimensional induced metric
$g_{\mu\nu}$, trace of the extrinsic curvature $K_{\mu\nu}$, the
brane-tension, and the matter Lagrangian on the $(\pm)$-branes,
respectively. Assuming {\boldmath{$Z$}}$_2$ symmetry across each
brane, the junction conditions to be imposed at each brane are
\begin{eqnarray}
K_{\pm\mu\nu} =r_c \left[-\kappa^2 \left(T^{(\pm)}_{\mu\nu} -\frac{1}{3}T^{(\pm)} g_{\mu\nu} \right)
                   + \left( G_{\pm \mu\nu} -\frac{1}{3}G_\pm g_{\mu\nu}
                              \right)
           \right], \label{junction}
\end{eqnarray}
with $r_c := \kappa^2 /2 \kappa_4^2$.

We assume vacuum without any matter field, $L_{m\sigma}=0$,
as an unperturbed state.
The five dimensional metric is given by
\begin{eqnarray}
ds^2 = dy^2 + a^2(y) \gamma_{\mu\nu} dx^\mu dx^\nu ,
\label{unpM}
\end{eqnarray}
with two boundary four dimensional de Sitter ($\pm$)-branes,
where $\gamma_{\mu\nu}$ is the four dimensional
de Sitter metric with unit curvature
radius and the warp factor $a(y)$ is given by
\begin{equation}
 a(y)=a_+ + y.
\end{equation}
The range of $y$-coordinate is from $0$ to $(a_--a_+)$.
The positions of $(+)$ and $(-)$-branes are at $y =y_+ :=0$ and
at $y=y_-:=a_--a_+$, respectively.
In the present model, the value of the warp factor
on the brane is related to
the Hubble parameters 
evaluated on each brane $H_\pm$ as
$a_{\pm}=1/H_\pm$.
Notice that the bulk geometry of the above solution is nothing but
the five dimensional
Minkowski spacetime written in the spherical Rindler coordinates.
From the junction conditions (\ref{junction}),
the Hubble parameters 
on the respective branes are related to the tension
of the branes as
\begin{eqnarray}
\pm H_\pm = r_c H^2_\pm - \frac{\kappa_4^2}{6} \tau_\pm , \qquad
 \mbox{with} \quad H_+ >H_-\geq 0 .
\end{eqnarray}

\subsection{Perturbation equations}

Now we study perturbations around the background mentioned above.
Besides the transverse-traceless conditions, one can impose
the conditions of vanishing $\{yy\}$ and $\{y\mu\}$-components 
of metric perturbations.
Then the perturbed metric is given by
\begin{eqnarray}
ds^2 = dy^2 + \left( a^2 \gamma_{\mu\nu} + h^{(TT)}_{\mu\nu} \right) dx^\mu
 dx^\nu,
\end{eqnarray}
with
\begin{eqnarray}
 \nabla^\mu h^{(TT)}_{\mu\nu} =0, \quad h^{(TT)\mu}_{~~~~ \mu}=0,
\end{eqnarray}
where $\nabla_\mu$ is the covariant derivative operator
associated with $\tilde\gamma_{\mu\nu}:=a^2\gamma_{\mu\nu}$.
In raising or lowering Greek indices, we use
$\tilde\gamma_{\mu\nu}$.
The $\{\mu\nu\}$-components 
of the Einstein equations in the bulk are
\begin{eqnarray}
a^2 h''^{(TT)}_{\mu\nu} -2h^{(TT)}_{\mu\nu} = -( \Box^{(4)}  -2 )h^{(TT)}_{\mu\nu},
\label{Einstein}
\end{eqnarray}
where a prime ``$~'~$'' denotes a partial differentiation with respect
to $y$,
and $\Box^{(4)}:=\gamma^{\mu\nu}\nabla_{\mu}\nabla_\nu$.

In this gauge, branes do not in general reside at fixed values of
$y$.  We denote the locations of the branes by $y = y_\pm +
\hat\xi^{(5)}_\pm (x^\mu)$. To study the boundary conditions, it
is convenient to introduce two sets of new ``Gaussian normal''
coordinates associated with respective branes. Here ``Gaussian
normal'' means coordinates such that $\{y\mu\}$-components vanish
and the location of each brane is specified by a constant
$y$-surface. Notice that the ``Gaussian normal'' coordinates
associated with $(+)$-brane in general differ from those
associated with $(-)$-brane. In these coordinates the junction
conditions imposed on the branes is simple.  We associate an
over-bar ``$\,\bar{~}\,$'' with the perturbation variables in
these coordinates.

The junction conditions for metric perturbations in the
``Gaussian normal'' coordinates are
\begin{eqnarray}
\pm \bar h^{(\pm)}_{\mu\nu}{}' \mp 2 H_\pm \bar h^{(\pm)}_{\mu\nu}
= - \kappa^2\left[ T^{(\pm )}_{\mu\nu}
                 - \frac{1}{3}\tilde\gamma_{\mu\nu} T^{(\pm )
       }\right]
      + 2r_c\left[
  X^{(\pm)}_{\mu\nu}-{1\over 3}\tilde\gamma_{\mu\nu}X^{(\pm)}\right],
\label{hjunctionA}
\end{eqnarray}
with
\begin{eqnarray}
&& X^{(\pm)}_{\mu\nu} := -\frac{1}{2} \left(a^{-2}\Box^{(4)}
     \bar h^{(\pm)}_{\mu\nu} -
    \nabla_\mu \nabla_\alpha \bar h^{(\pm)\alpha}_{\,~~ \nu}
        - \nabla_\nu \nabla_\alpha \bar h^{(\pm)\alpha}_{\,~~ \mu}
     + \nabla_\mu \nabla_\nu \bar h^{(\pm)}
                  \right) \nonumber \\
&& \qquad\qquad\qquad
 -\frac{1}{2}\tilde\gamma_{\mu\nu} \left(\nabla_\alpha \nabla _\beta  \bar h_{(\pm)}^{\alpha \beta }
 - a^{-2} \Box ^{(4)} \bar h^{(\pm)} \right)+ a^{-2}  \left( \bar h^{(\pm)}_{\mu\nu}
 + \frac{1}{2} \tilde\gamma_{\mu\nu} \bar h^{(\pm)} \right).
\label{XdefA}
\end{eqnarray}
Here
$T^{(\pm )}:=\tilde \gamma^{\mu\nu}T^{(\pm)}_{\mu\nu}$ and
$X^{(\pm)}$ is defined in the same manner.
The generators of the
gauge transformation from the 
``Gaussian normal'' coordinates to the Newton gauge
are given by
$$\xi_{(\pm)}^y:=y-\bar y
=\hat\xi^y_{(\pm)}(x^\mu),
$$
and
$$\xi_{(\pm)}^\mu:=x^\mu-\bar x^\mu={1\over a}\gamma^{\mu\nu}
\partial_\nu \hat\xi^y_{(\pm)}(x^\rho)
 +\hat\xi^\mu_{(\pm)}(x^\rho).
$$
By this gauge transformation, the metric perturbations
transform as
\begin{eqnarray}
\bar h^{(\pm)}_{\mu\nu} (y) = h^{(TT)}_{\mu\nu} (y)
 - 2a \left(1-a H_\pm\right)\nabla_\mu\nabla_\nu
 \hat\xi^y_{(\pm)}
 - 2a \gamma_{\mu\nu}\hat\xi^y_{(\pm)}(x^\rho),
\label{transformjunctionA}
\end{eqnarray}
where we have chosen $\hat\xi^\mu_{(\pm)}$ so that the second term
in (\ref{transformjunctionA}) vanishes on the brane. Substituting
(\ref{transformjunctionA}) into the traceless part of
Eq.~(\ref{hjunctionA}), the perturbed junction conditions for
$h^{(TT)}_{\mu\nu}$ imposed at $y=y_\pm$ becomes
\begin{eqnarray}
&&\pm h'^{(TT)}_{\mu\nu} \mp 2 H_\pm h^{(TT)}_{\mu\nu}
+ r_c H^2_\pm \left( \Box^{(4)}-2\right)h^{(TT)}_{\mu\nu}
  = -\kappa^2 \Sigma^{(\pm)}_{\mu\nu},
\end{eqnarray}
with
\begin{eqnarray}
\Sigma^{(\pm )} _{\mu\nu} = \left( T^{(\pm )}_ {\mu\nu}
     -\frac{1}{4} \tilde\gamma^{\pm}_{\mu\nu} T^{(\pm )}
               \right)
  \pm \frac{2}{\kappa^2}(1 \mp 2H_\pm r_c)
    \left( \nabla_\mu  \nabla _\nu -\frac{1}{4}
          \gamma_{\mu\nu} \Box ^{(4)}
    \right)\hat\xi^y_{(\pm)},
\end{eqnarray}
where $\tilde\gamma^+_{\mu\nu}:=\tilde\gamma_{\mu\nu}|_{y=y_+}$.
Combining the bulk equations and the junction conditions, we
finally obtain
\begin{eqnarray}
\left[\hat L^{(TT)}
   + \frac{ \Box^{(4)}\! -2  }{a^2}
  \right]
 h^{(TT)}_{\mu\nu}(y) = \sum_{\sigma=\pm}\left(-2\kappa^2
 \Sigma^{(\sigma)}_{\mu\nu}
      -2r_c H^2_\sigma
    \left(\Box^{(4)}\! -2\right)h^{(TT)}_{\mu\nu}(y_\sigma) \right)
   \delta(y-y_\sigma ),
\label{tensormasterA}
\end{eqnarray}
where
\begin{equation}
\hat L^{(TT)}:={1\over a^2}\partial_y a^4\partial_y{1\over a^2}.
\label{LTTdef}
\end{equation}
The trace part of Eq.~(\ref{hjunctionA})
gives
\begin{eqnarray}
\left(\Box ^{(4)}\! +4\right) \hat\xi^y_{(\pm)} =
  \pm \frac{\kappa^2}{6 H_\pm^2(1\mp 2r_c H_\pm)} T^{(\pm )}.
\label{xieqA}
\end{eqnarray}
The source $\Sigma^{(\pm)}_{\mu\nu}$ should also satisfy the transverse
conditions,
which can be shown to be identical to the condition~(\ref{xieqA})
by using the identity,
\begin{equation}
 \nabla^\mu
\left(\nabla_\mu \nabla_\nu -\frac{1}{4}
          \gamma_{\mu\nu} \Box^{(4)} \right)Y
\equiv
 {3\over 4} \nabla_\nu
\left(\Box ^{(4)}\!+4\right)Y,
\label{critical}
\end{equation}
which holds for an arbitrary scalar function $Y$.

\subsection{Solution with source}

Equations (\ref{tensormasterA}) are separable.
We define a complete set of functions of $y$ by (real-valued)
eigen-functions of the eigenvalue equation associated with
(\ref{tensormasterA}),
\begin{eqnarray}
\left[ \frac{m_i^2}{a^2}
   \left(1+2r_c\sum_{\sigma=\pm}\delta(y-y_\sigma)\right)\right]
           u_i(y) = -
          \hat L^{(TT)}
                 u_i(y),
\label{u}
\end{eqnarray}
where $m_i^2$ is the eigenvalue, which we find corresponds to
$\Box^{(4)}-2$ in comparison with Eq.~(\ref{tensormasterA}). Then,
thus defined eigen-functions are mutually orthogonal with respect
to the inner product defined by
\begin{eqnarray}
(u_i,u_{j})^{(TT)}:= \oint \frac{d\tilde y}{a^2}\left( 1 + 2 r_c
  \sum_{\sigma=\pm } \delta(y -y_\sigma)\right) u_i (y) u_{j} (y),
\label{innerproduct}
\end{eqnarray}
where the integral is taken over a twofold covering coordinate
$\tilde y$ which ranges from $-(y_--y_+)$ to $+(y_--y_+)$ and is
related to $y$ by $y=|\tilde y|$. Both ends of the range of
$\tilde y$ are identified. By definition, the norm
$(u_i,u_i)^{(TT)}$ is positive definite. We normalize the
eigen-functions so that they satisfy
$(u_i,u_{j})^{(TT)}=\delta_{ij}$.

Expanding $h^{(TT)}_{\mu\nu}$ in terms of these eigen-functions as
\begin{eqnarray}
h^{(TT)}_{\mu\nu} = \sum_j h_{\mu\nu}^{(j)} (x^\mu) u_j (y),
\end{eqnarray}
the equations of motion (\ref{tensormasterA}) reduce to
\begin{eqnarray}
{1\over a^2}\sum_j \left(\Box ^{(4)}\! -2 -m_j^2\right)\left(1+2r_c
  \sum_{\sigma=\pm} \delta(y-y_\sigma) \right)
  h_{\mu\nu}^{(j)}(x^\mu) u_j(y)
   = -2\kappa^2 \Sigma^{(+)}_{\mu\nu} \delta(y-y_+),
\end{eqnarray}
where we have assumed that the source is only on the $(+)$-brane.
Operating $(\oint d\tilde y\, u_i (y)\times)$ on both sides of the
above equations, we obtain
\begin{eqnarray}
    (\Box ^{(4)}\! -2 -m_i^2) h_{\mu\nu}^{(i)}(x^\mu)
    = -2\kappa^2 \Sigma^{(+)}_{\mu\nu} u_i (y_+).
\end{eqnarray}
Hence, the solution for
$h_{\mu\nu}$ becomes
\begin{equation}
h_{\mu\nu}^{(TT)} (y)
 = -2\kappa^2 \sum_i
    { u_i (y_+) u_i(y) \over \Box ^{(4)}\! -2 -m_i^2}
    \Sigma^{(+)}_{\mu\nu}.
\label{eq:solhTTA}
\end{equation}

From (\ref{transformjunctionA}) with the aid of
(\ref{xieqA}) and (\ref{eq:solhTTA}),
the induced metric on the $(+)$-brane for
the solution with source is
given by
\begin{eqnarray}
 \bar h^{(+)}_{\mu\nu}
 & = & -2\kappa^2\sum_i {u_i^2(y_+)\over \Box^{(4)}-2-m_i^2}
     \left(T_{\mu\nu}^{(+)}-{1\over 4}\tilde\gamma_{\mu\nu}T^{(+)}
      +{H_+^{-2}\over 3
            (m_i^2-2)}\left(\nabla_\mu\nabla_\nu
        -{1\over 4}\gamma_{\mu\nu}\Box^{(4)}
        \right)T^{(+)}\right)\cr
   && + \frac{\kappa^2}{6H_+^2}\gamma_{\mu\nu}
     \left({2\over H_+} (2r_cH_+ -1)^{-1}
        -
     \left( \sum_i \frac{u_i^2(y_+)}{m_i^2 -2}
         \right)
    {\Box^{(4)}}
       \right){1\over \Box^{(4)}+4}T^{(+)},
\label{soltensorA}
\end{eqnarray}
where we have neglected terms which can be erased by a gauge
transformation,
and we have used the relation
\begin{equation}
{1\over \Box^{(4)}\!-2-m^2}\left(\nabla_\mu\nabla_\nu
   -{1\over 4}\gamma_{\mu\nu}\Box^{(4)}\right)Y
=  \left(\nabla_\mu\nabla_\nu
   -{1\over 4}\gamma_{\mu\nu}\Box^{(4)}\right)
   {1\over \Box^{(4)}\!-m^2+6}Y,
\end{equation}
which holds
for an arbitrary scalar function $Y$.
This relation is equivalent to the identity
\begin{eqnarray}
\left(\nabla_\mu\nabla_\nu\Box^{(4)}
      -\Box^{(4)}\nabla_\mu\nabla_\nu\right)Y
\equiv  -8 \left(\nabla_\mu\nabla_\nu
   -{1\over 4}\gamma_{\mu\nu}\Box^{(4)}\right) Y.
\label{comutation}
\end{eqnarray}
Note that we have also used the relation
\begin{equation}
\frac{1}{(\Box^{(4)} -2-m_i^2)(\Box^{(4)}-4)}
= \frac{1}{m_i^2-2} \left( \frac{1}{\Box^{(4)} -2-m_i^2}
-\frac{1}{\Box^{(4)}-4} \right),
\label{spin02}
\end{equation}
which cannot be applied when $m_i^2 =2$.

We stress that when $m_i^2 =2$, there is no pathology present in
the model,
in contrast to the 
Fierz-Pauli case. In the 
Fierz-Pauli model, 
we cannot put the matter with $T \neq 0$ if the squared mass is
set to $2H^2$, because the equation of motion implies that $T=0$
(see Eq.~(\ref{APtrace}) in the Appendix). However, in the DGP
brane world model there is no such a restriction.
When the DGP brane world 
action is reduced to the form of Eq.~(\ref{soltensorA}), there
appears to be a similar pathology. However, this naive guess is
wrong since Eq.~(\ref{soltensorA}) is derived by using
Eq.~(\ref{spin02}), which is not correct when $m_i^2 =2$. When we
go back to more basic equations (\ref{eq:solhTTA}) and
(\ref{xieqA}), it is easy to find that the matter with $T \neq 0$
can be
consistently 
introduced into the DGP brane world 
even if $m_i^2 =2$.
The seemingly pathological expression (\ref{soltensorA}) at $m_i^2=2$
is due to the decomposition
to spin-2 and spin-0 perturbations (\ref{spin02}).

\subsection{Effective action}
\label{sec:effzero} {}From the expression (\ref{soltensorA}), we
find that the physical gravitational degrees of freedom are
composed of a Kaluza-Klein tower for the transverse-traceless part
and a single mode corresponding to the trace part in $\bar
h^{(+)}_{\mu\nu}$. We denote them by $\bar h^{(i)}_{\mu\nu}$ and
$s$, respectively. The induced metric perturbation is given by
\begin{eqnarray*}
\bar h^{(+)}_{\mu\nu}=\sum_i \bar h^{(i)}_{\mu\nu}+
  {1\over 4}\tilde\gamma^+_{\mu\nu} s.
\end{eqnarray*}
The kinetic term of the effective action is written as
\begin{eqnarray}
S_{kin} &=& \sum_i \alpha _i \int d^4x \sqrt{-\tilde\gamma_+}\,
      \bar h^{(i)\mu\nu} (\Box -2-m_i^2 ) \bar h_{\mu\nu}^{(i)}, \cr
 && + \beta \int d^4x \sqrt{-\tilde \gamma_+}\, s
     (\Box + 4) s,
\end{eqnarray}
where the indices are raised or lowered by using
$\tilde\gamma^+_{\mu\nu}$.
On the other hand,
the action of the matter localized on the $(+)$-brane,
expanded in powers of $\bar h^{(+)}_{\mu\nu}$, is written as
\begin{eqnarray}
S^{(+)}_{matter}
 =(\mbox{0-th order of } \bar {\bf h})
   &+& \sum_i \frac{1}{2}\int d^4 x \sqrt{-\tilde\gamma_+}\,
       T_{(+)}^{\mu\nu}
     \bar h_{\mu\nu}^{(i)} \nonumber \\
     &+& {1\over 8} \int d^4x \sqrt{-\tilde\gamma_+}\, T^{(+)}
      s              +O(\bar {\bf h}^2).
\end{eqnarray}

Comparing the equations of motion derived from the 
variation of $S_{kin}+S_{matter}$ with Eq.~(\ref{soltensorA}), we
can fix the coefficients, $\alpha_i$ and $\beta$. Here a little
subtlety arises when we take variation with respect to $\bar
h^{(i)}_{\mu\nu}$, since this variable is constrained to be
transverse-traceless. To avoid complication, here we simply
consider traceless source; $T^{(+)}=0$, deferring detailed
discussion to Appendix. Then, we find
\begin{eqnarray}
\alpha _{i} = {1\over 8 \kappa^2 u_i^2(y_+)} .
\end{eqnarray}
As for the trace part, by comparing
simply the poles at $\Box^{(4)}+4=0$, we obtain
\begin{eqnarray}
\beta= -\frac{3 H_+^2}{64\kappa^2}
     \left({1\over H_+} (2r_cH_+ -1)^{-1}
        +2
     \left( \sum_i \frac{u^2_i(y_+)}{m_i^2 -2}
         \right)
       \right)^{-1}.
\label{betafirst}
\end{eqnarray}

\begin{figure}[tb]
  \begin{center}
    \includegraphics[keepaspectratio=true,height=60mm]{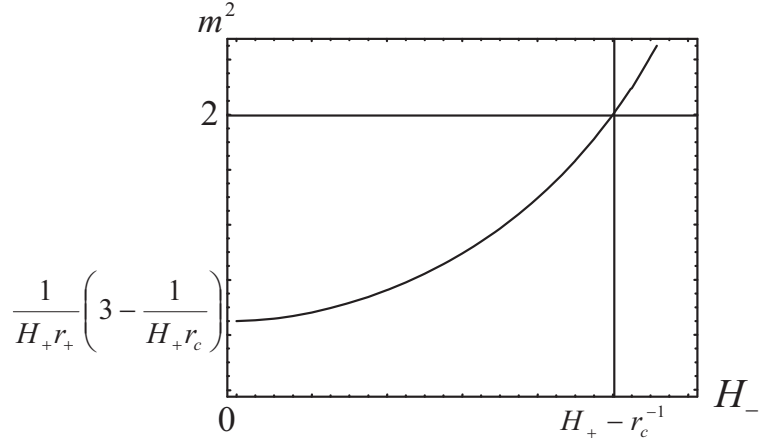}
  \end{center}
  \caption{$m^2$ as a function of $H_-$}
  \label{fig:mm.eps}
\end{figure}

The remaining tasks are to find the mass spectrum
of the 
eigenvalue equation (\ref{u}) and to evaluate $\beta$ in
(\ref{betafirst}) more explicitly. We begin with investigating the
mass spectrum. We are interested in the modes whose mass
eigenvalue $m^2$ is less than 2, since those modes contain a ghost
(except for the case with $m^2$ exactly equal to 0). 
In the bulk Eq.~(\ref{u}) can be solved as
\begin{eqnarray}
u_j = B_+ y^{\nu_+} + B_- y^{\nu_-},
\end{eqnarray}
where $B_\pm$ are some constants and
\begin{eqnarray}
\nu_\pm = 1/2 \pm \sqrt{ 9/4 -m_j^2}.
\end{eqnarray}
From the conditions at $y=y_\pm$,
we obtain
\begin{eqnarray}
&&A \cdot \textbf{B} =0, \label{eq:f12} \qquad \textbf{B} := \left(
  \begin{array}{c}
     B_+  \\
     B _- \\
  \end{array}
\right) , \\
&&A \! := \! \left(
  \begin{array}{cc}
  \! \left( \! - \frac{3}{2} +m_j^2 H_+ r_c \! + \! \sqrt{\frac{9}{4}-m_j^2} \right) \! H_+^{-\nu_+}  & \! \left( \! - \frac{3}{2} +m_j^2 H_+ r_c \! - \! \sqrt{\frac{9}{4}-m_j^2} \right) \! H_+^{-\nu_-}    \\
 \! \left( \! - \frac{3}{2} -m_j^2 H_- r_c \! + \! \sqrt{\frac{9}{4}-m_j^2} \right) \! H_-^{-\nu_+}       & \! \left( \! - \frac{3}{2} -m_j^2 H_- r_c \! - \! \sqrt{\frac{9}{4}-m_j^2} \right) \! H_-^{-\nu_-}   \\
  \end{array}
\right). \qquad
\end{eqnarray}
To obtain non-trivial solutions ($\textbf{B} \ne 0$),
$\det A=0$  must be satisfied.
Figure~\ref{fig:mm.eps} shows the relation between $H_-$ and
$m^2$ determined by $\det A =0$.
Fig.~\ref{fig:mm.eps} shows that
no ghost mode appears if the relation
\begin{eqnarray}
H_- > H_s := H_+ -r_c^{-1}, \label{Hs}
\end{eqnarray}
is satisfied.
Otherwise, one ghost mode whose mass eigenvalue $m^2$
is smaller than 2 appears.

Next we turn to the issue of finding a more explicit form
for $\beta$. For this purpose, we solve
\begin{equation}
\left(\hat L^{(TT)} +{2\over a^2}\right)
   Q(y)+4r_c \sum_{\sigma=\pm}H_\sigma^2 Q(y)\delta(y-y_\sigma)
 =-J\delta(y-y_+),
\label{Qeq}
\end{equation}
in two different ways.
First we solve this equation by expanding $Q(y)$ in terms of $u_i(y)$ as
$$
Q(y)=\sum_{i=0}^\infty Q^{(i)}u_i(y).
$$
Substituting this expansion into the above equation (\ref{Qeq})
and using the orthonormal conditions of the eigen-functions, we
obtain
\begin{eqnarray*}
 Q(y)=\sum_{i=0}^\infty {u_i(y_+)u_i(y)\over m_i^2 -2} J.
\end{eqnarray*}

Another way is to solve Eq.~(\ref{Qeq}) more explicitly.
The bulk part is solved by
\begin{eqnarray}
Q(y) = C_{0} +C_{1} a(y),
\label{C0C1A}
\end{eqnarray}
and $C_{0}$ and $C_{1}$ are constants to be determined by the
conditions at $y=y_\pm$.
These conditions are
\begin{eqnarray}
&& 2H_+ \left( -1 +H_+ r_c \right)
C_{0} + \left( -1 +2 H_+ r_c \right) C_{1}
= -{1\over 2}J, \cr
&&2H_- \left(1 +H_- r_c \right) C_{0} + \left(1 +2 H_- r_c \right) C_{1}
= 0.
\end{eqnarray}
Solving these for $C_{0}$ and $C_{1}$, we obtain
\begin{eqnarray}
&&C_{0} =
\frac{-( 1+ 2H_- r_c )}
   {4H_+ (H_+ r_c-1 )( 1+ 2H_- r_c )
   - 4H_- (2 H_+ r_c -1)( 1+ H_- r_c )}J,\cr
&&C_{1} = \frac{H_{-} ( 1+ H_- r_c )}
   {2H_+ (H_+ r_c-1 )( 1+ 2H_- r_c )
   - 2H_- (2 H_+ r_c -1)( 1+ H_- r_c )}J.
\label{C01A}
\end{eqnarray}
Substituting (\ref{C01A}) into Eq.~(\ref{C0C1A}),
we obtain $Q(y)$.
Comparing the expressions for $Q(y_+)$ evaluated in
these two different ways,
we find
\begin{eqnarray}
\sum_{i=0}^{\infty}{u_i^2(y_+)\over m_i^2-2} =
   {1\over H_+}\frac{2H_-(1+H_- r_c)-H_+( 1+ 2H_- r_c )}
   {4H_+ (H_+ r_c-1 )( 1+ 2H_- r_c )
   - 4H_- (2 H_+ r_c -1)( 1+ H_- r_c )}.
\end{eqnarray}
Substituting this into Eq.~(\ref{betafirst}), we finally find
\begin{eqnarray}
\beta & = & \frac{3 H_+^2 (2H_+ r_c -1)}{64\kappa^2(1+2H_- r_c)}
   \left({H_+ (H_+ r_c-1 )(1+2H_- r_c )
   - H_- (2 H_+ r_c -1)( 1+ H_- r_c )}\right)\cr
 &=& \frac{3 H_+^2 (2H_+ r_c -1)^2}{64\kappa^2}
   \left({\hat H_+ (1+\hat H_+ r_c)\over (1+2\hat H_+ r_c )}
   - {H_- (1+ H_- r_c)\over ( 1+ 2H_- r_c )}\right),
\end{eqnarray}
where $\hat H_+ :=H_+-r_c^{-1}$. One can see $\beta$ changes its
signature at $H_-=\hat H_+$. When $H_->\hat H_+$ ($H_-<\hat H_+$),
$\beta$ is negative (positive). This means that only when the
spin-2 ghost modes are absent, the spin-0 mode becomes a ghost
mode. (See Eq.~(\ref{Hs}).) Consequently, this model always has a
ghost.

\section{Stabilization}
\label{Stabilization}

In the preceding section, we have seen that when the spin-2
excitations have no ghost, the spin-0 brane-bending excitation
becomes a ghost instead. Usually spin-2 excitations (i.e.,
transverse-traceless tensor modes) are completely decoupled from
spin-0 excitations at linear order.  However, we can construct a
transverse-traceless tensor from a spin-0 mode whose mass squared
is $-4H^2$. For a scalar function $\Phi$ that satisfies
$(\Box^{(4)}+4)\Phi=0$, we can compose a transverse-traceless
tensor as
\begin{eqnarray}
\Psi_{\mu\nu}
= \left( \nabla_{\mu} \nabla_{\nu}-{1\over 4}
\gamma_{\mu\nu} \Box \right) \Phi.
\end{eqnarray}
It is trivial to show that this tensor is traceless. The
transverse property, $\nabla^\mu\Psi_{\mu\nu}=0$, can be shown by
using Eq.~(\ref{critical}). $\Psi_{\mu\nu}$ is also an
eigen-function of $\Box^{(4)}$, but its eigenvalue differs from
that of $\Phi$, which is $-4$. By using the identity
(\ref{comutation}), the mass eigenvalue of the induced spin-2
perturbation $\Psi_{\mu\nu}$ for the operator $(\Box^{(4)}-2)$ is
found to be
\begin{eqnarray}
m^2 =2.
\end{eqnarray}
Usually, we identify $H_\pm^2 m^2$ as the mass of a rank-two
tensor field seen by observers on the $(\pm)$-brane. Hence, the
mass squared of this special mode is $2H_\pm^2$ in the language of
spin-2 perturbations.  On the other hand, the mass squared of the
corresponding spin-0 mode, which is the eigenvalue of
$H^2\Box^{(4)}$, is $-4H_\pm^2$. Here, we refer to the mass of
this mode as the critical mass. This mass agrees with the
threshold mass below which the spin-2 ghost appears. If the mass
of a spin-2 mode is smaller than this critical mass, the helicity
0 part of the mode becomes a ghost \cite{Higuchi}.

In the two-branes model discussed in the preceding section, the
unique spin-0 excitation mode has exactly this critical mass. On
the other hand, $m^2=2$ is the threshold spin-2 mass below which a
ghost appears. If we gradually decrease the separation of the two
branes, the ghost in spin-2 excitations eventually disappears. The
mode that was previously carrying a ghost satisfies $m^2=2$ at the
critical point. At this point this spin-2 mode can couple with the
spin-0 excitations as an exceptional case. After passing the
critical point, the ghost is transferred to the spin-0
excitations.

Then, what happens if we introduce a mechanism to stabilize the
separation of the branes?  Since the mass spectrum of spin-0
excitations changes \cite{TM}, the spin-0 mode with its mass
squared being $-4H^2$ will no longer exist. In fact, the smallest
mass squared must be large enough if the stabilization mechanism
works.  Then, again we consider the situation in which the brane
separation is gradually decreased. When the spin-2 ghost
disappears, we expect absence of the spin-0 mode at the critical
mass, which can exceptionally couple to spin-2 excitations.
Then there seems to be no reason why 
the spin-2 ghost is transferred to the spin-0 ghost
at the critical point.

However, by studying models stabilized by a bulk scalar field,
we will find below that our naive expectation is wrong.
What we will see is the following.
With a bulk scalar field, there appears an infinite tower
of Kaluza-Klein modes for spin-0 excitations.
The smallest value of the mass squared of spin-0 excitations
is not, in general, equal to $-4H^2$ as is expected.
However, at the critical point where the smallest mass eigenvalue
for spin-2 excitations 
$m^2$ crosses $2$,
one of the values of the mass squared of spin-0 excitations
also crosses the critical value $-4H^2$.
As a result,
a spin-0 mode whose mass squared
is smaller than $-4H^2$ arises 
if the smallest $m^2$ is greater than 2, and the spin-0 mode
becomes a ghost. We discuss how this happen in detail below.

\subsection{Background}

We introduce a bulk scalar field to stabilize the brane
separation \cite{TM}. The action is given by
\begin{eqnarray}
&&S = S_2 + S_s, \\
&&S_s := \int d^5x  \sqrt{-g} \left( -\frac{1}{2}g^{ab}\psi_{,a}
  \psi_{,b} -V_B (\psi )
- \sum_ {\sigma= \pm } V^{(\sigma)}(\psi)
\delta(y-y_\sigma) \right) .
\end{eqnarray}
By choosing the potential in the bulk $V_B (\psi)$ and the
potentials on the branes $V^{(\sigma)}(\psi)$ appropriately,
we can stabilize the brane separation.
In the following discussion we do not need the 
explicit form of the potential functions.
The unperturbed background configuration is similar to
the previous case.
The bulk metric is given by
(\ref{unpM}) and
the branes are located at a fixed value of $y(=y_{\pm})$
as before.
But the functional form of
the warp factor $a(y)$ is different.
The background bulk equations for the warp factor $a(y)$ and
the scalar field $\psi$ are
\begin{eqnarray}
&&\frac{a''}{a} - \left( \frac{a'}{a} \right) ^2
  = - \frac{\kappa^2}{3} \psi'{}^2 -\frac{1}{a^2} ,\\
&&\left( \frac{a'}{a} \right) ^2 = \frac{\kappa^2}{6} \left( \frac{1}{2}
\psi'{}^2 - V_B
 \right) + \frac{1}{a^2},\\
&&\psi'' + 4 \frac{a'}{a} \psi' - {\partial V_{B}\over \partial\psi} =0.
\end{eqnarray}

\subsection{Perturbations}


We use the 
``Newton gauge'', in which the spin-0 component
of the shear of the hypersurface
normal vector vanishes, following Ref.~\cite{TM}.
In this gauge, using the traceless part and $\{y\mu\}$-component
of the Einstein equations, we find that
perturbations of the metric and the scalar field are related as
\begin{eqnarray}
&&h_{yy} =2\phi, \quad h_{y\mu} =0,\cr
&&h_{\mu\nu} = h^{(TT)}_{\mu\nu} - \phi \tilde\gamma_{\mu\nu},\cr
&&\delta \psi = \frac{3}{2 \kappa^2 \psi'}\left[ \partial_y + 2
    {\cal H} \right]\phi,
\label{formofperturbation}
\end{eqnarray}
where $\tilde \gamma_{\mu\nu}:= a^2(y)\gamma_{\mu\nu}$, ${\cal
H}:=a'/a$ and $h_{\mu\nu}^{(TT)}$ is a tensor which satisfies
transverse-traceless conditions. Here, one remark is in order. For
modes at the critical mass there is a possible coupling between
spin-0 and spin-2 excitations. Recall that
the traceless tensor generated by acting a derivative operator 
from a scalar-type function
($[ \nabla_\mu \nabla_\nu - \frac{1}{4} \gamma _{\mu\nu} \Box^{(4)} ] \Phi$ )
is automatically transverse.
Due to this mixing, vanishing of
spin-0 component of the shear is slightly ambiguous.
Here we define the ``Newton gauge'' by the above form
of perturbations (\ref{formofperturbation}),
and therefore it actually can possess spin-0 component of
the shear for the mode with this critical eigenvalue.
In this sense, the gauge is not completely fixed.

In the present gauge, the bulk equations for $h^{(TT)}_{\mu\nu}$
become
\begin{eqnarray}
&&\left[ \hat L ^{(TT)} + \frac{1}{a^2} ( \Box ^{(4)}\! -2 )
   \right]
 h^{(TT)}_{\mu\nu} =0,
\end{eqnarray}
with $h^{(TT)}$ as defined in (\ref{LTTdef}).
While, the bulk equation for $\phi$ becomes
\begin{eqnarray}
&&\left[\hat L^{(\phi)}+{1\over \psi'{}^2}
   (\Box^{(4)}\!+4)
     \right] \phi =0,
\end{eqnarray}
with
\begin{eqnarray}
&&\hat L^{(\phi )} := a^2 \partial_y \frac{1}{a^2 \psi'{} ^2}
 \partial_y a^2 - \frac{2 \kappa^2}{3}a^2.
\end{eqnarray}


We first write down the junction conditions to impose
at the branes in the 
``Gaussian normal'' coordinates.
Perturbation variables with an over-bar ``$~\bar{}~$'' are
understood as those in the 
``Gaussian normal'' coordinates, as before. The junction
conditions for metric perturbations are
\begin{eqnarray}
\pm \left(\partial _{\bar y} -2{\cal H}\right)
   \bar h^{(\pm)}_{\mu\nu} = - \kappa^2\left[ T^{(\pm )}_{\mu\nu}
                 - \frac{1}{3}\tilde\gamma_{\mu\nu} T^{(\pm )}\right]
   \mp \frac{2\kappa^2}{3}\tilde\gamma_{\mu\nu}
           \psi' \delta\psi + 2r_c\left[
  X_{\mu\nu}-{1\over 3}\tilde\gamma_{\mu\nu}X\right],
\label{hjunction}
\end{eqnarray}
with $X_{\mu\nu}$ defined in (\ref{XdefA}).
While, the condition for the bulk scalar field is
\begin{eqnarray}
\pm 2 \delta {\bar \psi'} = V''^{(\pm)} (\psi ) \delta \bar \psi.
\label{scalarJ}
\end{eqnarray}

The generators of the
gauge transformation from the 
``Gaussian normal'' coordinates to
the Newton gauge take the form of
\begin{eqnarray}
&&\xi^y _{(\pm )} = \int^y \phi (y') dy' =
\int _{y_{\pm}} ^y \phi(y') dy' + \hat \xi^y_{(\pm )}(x^\mu) ,
\label{xiydef}\\
&&\xi_{(\pm)}^\nu
 = - \int ^y \tilde\gamma^{\mu\nu} (y') dy' \int ^{y'}\phi_{,\mu}(y'') dy'' \nonumber\\
&&\qquad
= -\int ^y_{y_{\pm}}\tilde\gamma^{\mu\nu} (y')
\left[\int_{y_{\pm}}^{y'} \phi_{,\mu}(y'')dy''
    +\hat\xi^y_{(\pm ),\mu}(x^\rho)\right]dy'
   + \hat \xi ^\nu _{(\pm )}(x^\rho).
\label{xinudef}
\end{eqnarray}
Under this gauge transformation, the perturbation variables
transform as
\begin{eqnarray}
&&\bar h^{(\pm)}_{\mu\nu} (y) = h_{\mu\nu} (y)
 + 2\nabla{}_{(\mu} \xi{}_{\nu)}{}^{\!\!\!\!\!\!(\pm)}
-2{\cal H}\tilde\gamma_{\mu\nu} (y) \xi^y_{(\pm)}, \\
&&\delta \bar \psi^{\pm} (y) = \delta \psi (y)
- \psi' \xi^y_{(\pm)}
+ \hat \xi^y_{(\pm )}  .
\label{transformjunction}
\end{eqnarray}

Substituting these relations into Eqs.~(\ref{hjunction})
and (\ref{scalarJ}), we obtain
the junction conditions in the 
Newtonian gauge. The conditions for the traceless part of metric
perturbations become
\begin{eqnarray}
\pm \left(\partial_y - 2{\cal H}_\pm \right) h^{(TT)}_{\mu\nu} =
 -\kappa^2 \Sigma_{\mu\nu}^{(\pm )}
  - r_c H_\pm^2({\Box ^{(4)}\! -2 })h^{(TT)}_{\mu\nu},
  \label{junTT}
\end{eqnarray}
with ${\cal H}_\pm:={\cal H}(y_\pm)$ and
\begin{eqnarray}
\Sigma^{(\pm )} _{\mu\nu} := \left( T^{(\pm)}_ {\mu\nu}
     -\frac{1}{4} T^{(\pm )} \tilde\gamma_{\mu\nu} \right)
  \pm \frac{2}{\kappa^2}
    \left( \nabla_\mu  \nabla _\nu -\frac{1}{4}
          \gamma_{\mu\nu} \Box^{(4)} \right)Z_{(\pm)}. 
\end{eqnarray}
On the other hand,
the condition for the trace part of metric perturbations becomes
\begin{eqnarray}
H_\pm^2({\Box ^{(4)}\! +4}) Z _{(\pm)} =
  \pm \frac{\kappa^2}{6} T^{(\pm )},
\label{xieq}
\end{eqnarray}
where
\begin{equation}
 Z_{(\pm)}=\left(1\mp 2r_c{\cal H}_\pm\right)\hat\xi_{(\pm )} ^y
       -r_c \phi(y_\pm).
\end{equation}
The junction condition for the scalar-type perturbation becomes
\begin{eqnarray}
\pm \frac{2\kappa^2}{3}\left( \delta \psi - \psi'
\left(1\mp 2r_c{\cal H}\right)^{-1}(Z_{(\pm)}+r_c\phi)
 \right)
 = \frac{ \epsilon^{(\pm )}}{a^2 \psi'}
    \left( \Box^{(4)}\! + 4 \right) \phi ,
\label{scalarjun}
\end{eqnarray}
at $y=y_\pm$ with
\begin{eqnarray}
\epsilon^{(\pm)} := \frac{2}{V ''^{(\pm )} \pm 2
  \psi''/\psi' }\quad .
\end{eqnarray}
In the following discussion we assume $\epsilon^{(\pm)}\geq 0$,
which includes the simplest case of the tight binding limit;
$V''{}^{(\pm)}\to \infty$.

Combining the bulk equations and the junction conditions, we
formally obtain the same equation as (\ref{tensormasterA})
for spin-2 excitations, and
\begin{eqnarray}
\left[\hat L^{(\phi )} + {1\over \psi'{}^2}( \Box^{(4)}\! + 4 )
\right] \phi
=  \sum_{\sigma =\pm} \left(\sigma\frac{4a^2 \kappa^2}{3}
    \left(1-\sigma 2r_c{\cal H}_\pm\right)^{-1}(Z_{(\sigma)}+r_c\phi)
    -{2\epsilon^{(\sigma)} \over \psi'{}^2}(\Box^{(4)}\! +4) \phi\right)
  \delta(y-y_\sigma),
\label{scalarmaster}
\end{eqnarray}
for spin-0 excitations.

\subsection{Solution with source}
\label{section:5.1}

As was done in \S\ref{twobrane}, we formally write down a solution
of perturbation equations for a given energy momentum tensor by
using the Green's function written in terms of eigen-functions of
$\hat L ^{(TT)}$ and $\hat L^{(\phi)}$. We also derive expressions
for the metric perturbations induced on the branes.


We begin with spin-2 excitations. We expand the metric
perturbation $h_{\mu\nu}^{(TT)}$ by using the eigen-functions of
the eigenvalue problem (\ref{u}), as before.  The calculation
closely parallels the previous case, although the functional form
of the warp factor $a$ is not the same. We obtain
(\ref{eq:solhTTA}), formally the same expression as before.

Spin-0 excitations can be solved in the same way. We define the
eigen-functions $v_i(y)$ by
\begin{eqnarray}
\frac{\mu_i^2 +4}{\psi'{}^2} \left( 1+ \sum _{\sigma = \pm } 2 \epsilon
   ^\sigma \delta(y- y_\sigma ) \right) v_i(y)
   = \left[- \hat L^{(\phi )} +\sum_{\sigma=\pm}\sigma{4r_c \kappa^2 a^2\over 3}
     \left(1-\sigma 2r_c{\cal H}_\pm\right)^{-1} \delta(y- y_\sigma ) \right] v_i (y),
\end{eqnarray}
and we define the inner product by
\begin{eqnarray}
 (v_i,v_j)^{(\phi)}:= \oint \frac{dy}{\psi'{}^2}
  \left(1+\sum_{\sigma=\pm }
  2\epsilon^\sigma \delta(y-y_\sigma)\right) v_i(y) v_j(y).
\end{eqnarray}
Under the assumption that $\epsilon^\sigma\geq 0$, it is trivial
to show that the norm of each eigenfunction is positive definite,
and hence they can be normalized as
$(v_i,v_j)^{(\phi)}=\delta_{ij}$. Using thus normalized
eigen-functions, the solution of Eq.~(\ref{scalarmaster}) is given
by
\begin{eqnarray}
\phi (y) =\frac{4\kappa^2}{3H_+^2} \left(1- 2r_c{\cal H_+}\right)^{-1}
    \sum_i {v_i (y_+) v_i(y) \over \Box^{(4)}\!-\mu_i^2} Z_{(+)}.
\label{scalarsol}
\end{eqnarray}

The induced metric on the brane for
the solution with the source is given by
\begin{eqnarray}
 \bar h_{\mu\nu}^{(+)} & = & h_{\mu\nu}^{(TT)}(y_+)
         -\tilde \gamma_{\mu\nu}
     \left(\phi(y_+)+2{\cal H}_+ \hat\xi^y_{(+)}\right),
\label{eq:results}
\end{eqnarray}
Here, transverse-traceless part $h_{\mu\nu}^{(TT)}$
takes the same form as given in Eq.~(\ref{eq:solhTTA}),
and it can be rewritten more explicitly as
\begin{eqnarray}
 h_{\mu\nu}^{(TT)}(y_+)
 & = & -2\kappa^2\sum_i {u_i^2(y_+)\over \Box^{(4)}-2-m_i^2}
     \left(T_{\mu\nu}^{(+)}-{1\over 4}\tilde\gamma_{\mu\nu}T^{(+)}
      +{H_+^{-2}\over 3
            (m_i^2-2)}\left(\nabla_\mu\nabla_\nu
        -{1\over 4}\gamma_{\mu\nu}\Box^{(4)}
        \right)T^{(+)}\right)\cr
   && + \frac{2\kappa^2}{3H_+^2}\left( \sum_i \frac{u^2_i(y_+)}{m_i^2 -2}
         \right) \left( \nabla_\mu \nabla_\nu
        -\frac{1}{4} \gamma_{\mu\nu} \Box^{(4)}
        \right)\frac{1}{\Box^{(4)}+4}T^{(+)}.
\label{soltensor}
\end{eqnarray}
While the trace part is computed as
\begin{eqnarray}
 \phi+2{\cal H}_+\hat\xi^y_{(+)}
   & = &
   {2\kappa^2\over 9\tilde H_+^4}\sum_i{v_i^2(y_+)\over \mu_i^2 +4}
      {1\over \Box^{(4)}-\mu_i^2}T^{(+)}\cr
&&    -{\kappa^2\over 3 \tilde H_+^2}\left(
       {2\kappa^2\over 3 \tilde H_+^2}\left(\sum_i{v_i^2(y_+)\over
                  \mu_i^2 +4}\right)+{a'\over a}
     \right)
      {1\over \Box^{(4)}+4}T^{(+)},
\label{solscalar}
\end{eqnarray}
where for brevity we have introduced
\begin{equation}
\tilde H_+^2=H_+^2\left(2r_c{\cal H}_+-1\right).
\end{equation}

The above expression for $\bar h_{\mu\nu}^{(+)}$ apparently
contains a pole at the critical mass eigenvalue corresponding to
$(\Box^{(4)}+4)^{-1}$. The term proportional to $\nabla_\mu
\nabla_\nu$ in Eq.~(\ref{soltensor}) is pure gauge. Hence, the
contribution from the pole at the critical mass eigenvalue is
proportional to
 $\gamma_{\mu\nu}(\Box^{(4)}+4)^{-1}T^{(+)}$.
Collecting such terms in $\bar h_{\mu\nu}^{(+)}$
coming both from the $TT$-part and
from the scalar-part, the coefficient becomes
\begin{eqnarray}
 {\kappa^2 \over 3 H_+^2} \left\{2
            \left(
              \sum_i {u_i^2(y_+)\over m_i^2-2}\right)
 + {1\over \tilde H_+^2}\left(
       {2\kappa^2\over 3 \tilde H_+^2}\left(\sum_i{v_i^2(y_+)\over
                  \mu_i^2 +4 }\right)+{\cal H}_+
     \right)\right\}.
\label{m4}
\end{eqnarray}
In \S\ref{sec:NoCMM}
we will prove an identity which shows that this coefficient
vanishes as a whole.

\subsection{No physical degrees of freedom at critical mass}
\label{sec:NoCMM}

As we mentioned earlier, our ``Newton gauge'' is not
a complete gauge fixing.
In this section, we will fist show that there is no physical mode at
the critical mass in general after completely
fixing the residual gauge.
Next, we will show that the value of the
expression (\ref{m4}) equals $0$.


The residual gauge degree of freedom of our ``Newton gauge'' exists 
at the critical mass, i.e., when
$\Box^{(4)}\!=-4$ for scalar-type perturbations
and $\Box^{(4)}\!=4$ for transverse-traceless tensor perturbations.
To see this, let's consider the gauge transformation
generated by the following infinitesimal coordinate
transformation:
\begin{equation}
 \xi^y={1\over 2}
     a^2\partial_y \left({X\over a^2}\right),\qquad
 \xi^\mu=-{1\over 2}\tilde\gamma^{\mu\nu} X_{,\nu}.
\end{equation}
This gauge transformation keeps the conditions $h_{y\mu}=0$. The
change of $\phi$ induced by this gauge transformation can be read
off the $\{yy\}$-component of the metric perturbation:
\begin{equation}
\delta \phi={1\over 2} \partial_y \left(
     a^2\partial_y \left({X\over a^2}\right)\right).
\label{eq:phi1}
\end{equation}
At the same time, the induced change of $\phi$ can be also read
off the trace part of the metric perturbation:
\begin{equation}
\delta \phi=-a' a\,\partial_y \left({X\over a^2}\right)-{X\over a^2},
\label{eq:phi2}
\end{equation}
where we have used $\Box^{(4)}X=-4X$. Equating these two
expressions, we obtain an equation for $X$, which is
\begin{equation}
 \hat L^{(TT)}X=-{2\over a^2}X.
\label{eqX}
\end{equation}
Gauge transformations with $X$ satisfying this equation preserve
the ``Newton gauge'' conditions.  Hence, such gauge
transformations represent the residual gauge degrees of freedom.

Notice that the above equations are identical to the equations for
the $TT$-part (\ref{u}) with $\Box^{(4)}\! =4$. The traceless part
generated by this gauge transformation is
\begin{equation}
\delta h^{(TT)}_{\mu\nu}=-\left(\nabla_\mu\nabla_\nu
   -{1\over 4}\gamma_{\mu\nu}\Box^{(4)}\right)X.  \label{hTTX}
\end{equation}
Therefore, Eq.~(\ref{eqX}) means that the $h^{(TT)}_{\mu\nu}$
induced by the residual gauge transformation satisfies the
perturbation equation consistently.
It is also straight forward to show 
that $\phi=\delta\phi$ given by
Eq.~(\ref{eq:phi1}), or equivalently by Eq.~(\ref{eq:phi2}),
satisfies the equation for the scalar-type perturbation with $\Box^{(4)}\! =-4$.
To conclude, solutions of the bulk perturbation equations for
both spin-2 and spin-0 parts 
at the critical mass are generated by a 
gauge transformation.

Next, we turn to the
the issue of boundary conditions. 
As we are interested in the gravitational degrees of freedom, 
we consider cases without matter fields: 
$T^{(\pm)} =0$. From Eq.~(\ref{xieq}), we find $(\Box ^{(4)} + 4)
Z_{(\pm)} =0$. Both spin-2 and spin-0 parts are sourced by
$Z_{(\pm)}$ and they have solutions for any given $Z_{(\pm)}$.
However, the spin-2 part of the solution, which is proportional to
$\left(\nabla_\mu\nabla_\nu
   -{1\over 4}\gamma_{\mu\nu}\Box^{(4)}\right)Z_{(\pm)}$,
can be erased by the residual gauge transformation mentioned
above. By the definition of residual gauge transformation, it does
not violate the gauge conditions necessary to derive the set of
perturbation equations (\ref{tensormasterA}) and
(\ref{scalarmaster}). Therefore the solution after this gauge
transformation still satisfies all the perturbation equations.
Once the $TT$-part is completely erased by the gauge
transformation, the junction conditions therefore imply that
$Z_{(\pm)}$ also vanishes after this gauge transformation. The
equation of motion for the scalar part also continues to hold
after the gauge transformation. But now, with vanishing
$Z_{(\pm)}$, the only consistent solution for $\phi$ is
$\phi\equiv 0$, except for models with the potentials $V_B$ and
$V^{(\pm)}$ being fine tuned.

For any given homogeneous solution of $Z_{(\pm)}$, we can
construct a solution for the bulk perturbations which solves all
the perturbation equations. However, what we have found is that
seemingly non-trivial solutions constructed in such a manner are
all pure gauge modes related to the residual gauge degrees of
freedom of the ``Newton gauge''. As we have anticipated, there is
no physical perturbation mode at the critical mass.

We found that by fixing the residual gauge, we can set $Z_{(\pm)}=0$
in the source-free case,
and therefore we succeeded in eliminating the mode at the
critical mass from the
physical spectrum.
Using this fact, we can prove
that the expression presented in (\ref{m4}) vanishes as follows:
By making a residual gauge transformation to the unperturbed
background,
we can set $Z_{(+)}=Z_0$ 
for such $Z_0$ that satisfies
$(\Box^{(4)} +4 ) Z_0 =0$.
After the gauge transformation,
we obtain a pure gauge mode for
$h^{(TT)}_{\mu\nu}$ and $\phi$ at the critical mass.
From Eqs.~(\ref{eq:phi2}) and (\ref{hTTX}),
this gauge mode is given explicitly by
\begin{eqnarray}
&&h^{(TT)}_{c\,\mu\nu}=-\left(\nabla_\mu\nabla_\nu
   -{1\over 4}\gamma_{\mu\nu}\Box^{(4)}\right)X,
\label{hcTT}\\
&&\phi_c=-a' a\,\partial_y \left({X\over a^2}\right)-{X\over a^2}, \label{phic}
\end{eqnarray}
where a subscript $c$ indicates the mode at the critical mass.
Since it is a pure gauge mode,
the above mode should satisfy the junction conditions
(\ref{scalarjun}) and (\ref{junTT})
with $Z_{(+)}=Z_0$ 
and $T^{(\pm)}_{\mu\nu}=0$.
Both conditions give the same boundary condition for $X$.
Substituting (\ref{hcTT}) into (\ref{junTT}) at $y=y_+$,
we obtain
\begin{eqnarray}
\left. a^2 \partial_y \left(\frac{X}{a^2} \right)\right\vert_{y=y+}
   = 2Z_0 - \left. 2 r_c H_+^2{X}\right\vert_{y=y+}.
\end{eqnarray}
Substituting this into Eq.~(\ref{phic}) to eliminate
$\partial_y X$, we get
\begin{eqnarray}
\phi_c(y_+) = -2 {\cal H}_+ Z_0 - H_+^2
    \left( 1 - 2r_c {\cal H}_+ \right)X.
\label{phyc2}
\end{eqnarray}

On the other hand, from Eqs.~(\ref{eq:solhTTA}) and
(\ref{scalarsol}) with the aid of (\ref{hcTT}),
we have
\begin{eqnarray}
X = 4 \sum_i \frac{u_i^2 (y_+)}{2-m_i^2} Z_0, \qquad \mbox{and} \qquad
\phi_c = \frac{4\kappa^2}{3 \tilde H^2_+} \sum_i \frac{v^2_i (y_+)}{4+ \mu^2_i}Z_0.\label{phyc1}
\end{eqnarray}
Combining Eqs.~(\ref{phyc2}) and (\ref{phyc1}),
we finally obtain
\begin{eqnarray}
2\left(\sum_i {u_i^2(y_+)\over m_i^2-2}\right)
 + {1\over \tilde H_+^2}\left(
       {2\kappa^2\over 3 \tilde H_+^2}\left(\sum_i{v_i^2(y_+)\over
                  \mu_i^2 +4 }\right)+{\cal H}_+
     \right)=0 .
\label{mode4}
\end{eqnarray}
Namely, the coefficient of the pole at the critical mass,
(\ref{m4}), vanishes.

Although Eqs.~(\ref{soltensor}) and (\ref{solscalar}) apparently
contain poles at the critical mass, we find that they completely
cancel each other. This is consistent with the conclusion obtained
earlier in this subsection that there is no physical mode at the
critical mass.

\subsection{Effective action and ghost conditions}

Applying the same method used in \S\ref{sec:effzero}, we now
derive the four dimensional effective action to the quadratic
order for the present case. From the derived effective action, we
will identify the conditions for the presence of ghost.

We denote the Kaluza-Klein towers of the transverse-traceless
part and the trace part of the metric perturbation induced
on the $(+)$-brane by $\bar h^{(i)}_{\mu\nu}$
and $\bar h^{(j)}$, respectively.
The induced metric perturbation is given by
\begin{eqnarray*}
\bar h_{\mu\nu}=\sum_i \bar h^{(i)}_{\mu\nu}+
  {1\over 4}\tilde\gamma_{\mu\nu}\sum_j \bar h^{(j)}.
\end{eqnarray*}
Then, the kinetic term of the effective action is
\begin{eqnarray}
S_{kin} &=& \sum_i \alpha _i \int d^4x \sqrt{-g}\,
   \bar h^{(i)\mu\nu} (\Box -2-m_i^2 ) \bar h_{\mu\nu}^{(i)} \cr
 &&\qquad + \sum_j \beta_j \int d^4x \sqrt{-g}\, \bar h^{(j)}
     (\Box - \mu_j^2) \bar h^{(j)},
\end{eqnarray}
and the action of the matter localized on the $(+)$-brane
is
\begin{eqnarray}
S^{(+)}_{matter}
 =(\mbox{0-th order of } \bar {\bf h})
   &\!\!+\!\!& \sum_i \frac{1}{2}\int d^4 x \sqrt{-g} T_{(+)}^{\mu\nu}
     \bar h_{\mu\nu}^{(i)} \nonumber \\
     &\!\!+\!\!& \sum_j {1\over 8} \int d^4x \sqrt{-g} T^{(+)}
   \bar h^{(j)}
              +O(\bar {\bf h}^2).
\end{eqnarray}

Comparing the equations of motion derived from the 
variation of $S_{kin}+S_{matter}$ with Eqs.~(\ref{soltensor}) and
(\ref{solscalar}),
we find
\begin{eqnarray}
\alpha _{i} = {1\over 8 \kappa^2 u_i^2(y_+)}, \qquad
\beta_j = {9 \tilde H_+ ^4 (\mu^2_i +4)\over 128 \kappa^2 v_i^2(y_+)}.
\end{eqnarray}
All coefficients of the spin-2 component, $\alpha_i$,
are always positive.
However, when the mass eigenvalue is in the range $0<m^2_i<2$,
such a mode contains a ghost in its helicity zero component.
In contrast, the coefficients of spin-0 component, $\beta_i$,
becomes negative when $\mu_i^2 <-4$.
In this case this mode becomes a ghost.
Hence, to realize a ghost-free model,
all masses of spin-2 modes and spin-0 modes must
satisfy $m^2_i>2$ and $\mu_i^2>-4$.
However, it turns out to be impossible that both
conditions are satisfied simultaneously.

In order to show this, we consider continuous deformation of the
model potentials of the bulk scalar field. When the smallest
eigenvalue of spin-2 excitations, $m^2_0$, approaches 2 from
below, ${u_0^2(y_+)/(m^2_0 -2)}$ in Eq.~(\ref{mode4}) diverges.
Since this equality is always satisfied, this divergence must be
compensated by the other terms. The possibility is that one of the
eigenvalues of spin-0 excitations $\mu_j^2$ approaches $-4$ from
above. Therefore, when the value of $m^2_0$ exceeds 2, at least
one of the values, $\mu^2_j$, must be smaller than $-4$. Hence, we
cannot simultaneously satisfy $m_0^2>2$ and $\mu_0^2>-4$ by a
continuous deformation.  All models specified by given bulk scalar
field potentials are connected with each other by continuous
deformations.  Thus it is proved that we cannot realize a model
free from a
ghost within the two-brane extension of the 
DGP model with a bulk scalar field.

In the above discussion 
we have introduced the four dimensional effective action just
from the notion of the equations of motion.
Homogeneous equations of motion are sufficient to fix the form of the effective 
action except for  the overall normalization of each mode. However, to 
determine if there is 
a ghost or not, the overall normalization is  crucial . 
We have determined this overall normalization factor
by investigating the coupling to the matter energy momentum tensor.

There are alternative ways to fix this overall normalization for each mode
without looking at the coupling to matter fields.
We could have derived a four dimensional effective action directly
from the original five dimensional action, although we feel that the approach
adopted in this paper is slightly easier. 
One short cut to fix this normalization is 
just to look at the bulk part of the five dimensional reduced action 
written down in terms of the master variable.
To obtain the five dimensional action, we can use double Wick rotation:
$y\to iy$ and $t \to -it$, which transforms
the four dimensional de Sitter space into an Euclidean four sphere.
Then, the discussion in the bulk is completely parallel to
the cosmological perturbation
in the (five dimensional) closed FRW universe,
and we can make use of the knowledge about the effective action
in the context of the cosmological perturbation in the FRW
universe\cite{open}.
Since our action should be obtained by the double Wick rotation from this
action, one can easily read the correct form of the reduced action 
including the overall normalization.
By doing so, we will find that when and only when $\mu^2_i< -4$,
the overall normalization factor becomes negative for
spin-0 excitations.
For spin-2 excitations, the overall factor is always positive
definite. This gives an another route to reach
the same conclusion that we have obtained in this section.
\section{Summary}
\label{Summary}

In the present paper, we have attempted to construct a ghost-free model
by modifying the self-accelerating 
branch of the DGP brane world
scenario, in which the 
spin-2 sector drives the accelerated expansion of the universe. We
first tried adding an additional boundary brane in
\S\ref{twobrane}, and then stabilizing the brane separation by
introducing a bulk scalar field in \S\ref{Stabilization}. In both
cases, we found that the ghost excitation survives. As we have
already mentioned in Introduction, this is caused by the
degeneracy between spin-2 and spin-0 modes at the critical mass.
This degeneracy enables the ghost property to transmute between
these two normally decoupled modes. But this degeneracy occurs
only at this special mass.

In the case without stabilization, the fluctuation of brane
separation gives the spin-0 mode. Hence, if we fix the brane
separation, the mass spectrum of spin-0 excitations will be
significantly modified. We introduced Goldberger-Wise
mechanism~\cite{GW} into our two-branes model. Then, in general
the spin-0 spectrum does not have a mode at the critical mass.
However, when the spin-2 mode crosses the critical mass, we found
that the one of the mass of spin-0 mode also crosses the critical
value, and it turns into a ghost.
We have explained why we cannot avoid this exchange of a 
ghost degree of freedom in detail in the text.
The point is as follows: If we compute the effective action for
the matter field obtained after integrating out the gravitational
degrees of freedom, the spin-2 contribution unexpectedly gives the
term depending on the trace of the matter energy momentum tensor.
This is because the spin-2 perturbation is sourced by the brane
bending, which is sourced by the trace of the energy momentum
tensor. Although this brane bending does not appear in the
physical spectrum of the model once stabilization is introduced,
its mass still corresponds to the critical mass. As a result, when
one of the masses of the spin-2 excitations crosses the critical
value, the pole of the brane bending resonates with the pole of
this crossing mode. This leads to the divergence of the effective
action of the matter field. Physically, however, any divergence in
the effective action is not expected. This divergence is cancelled
by the contribution from the spin-0 mode. But such cancellation is
possible only if the mass spectrum of spin-0 modes also crosses
the critical mass at the same time.

It is instructive to emphasize the difference of
the behaviour of spin-2 perturbations in the DGP model
compared to the massive Fierz-Pauli model. 
In the massive Fierz-Pauli model, 
a pathology appears if the mass becomes $2H^2$ because
the equation of motion implies that $T=0$.
In the DGP brane world, 
matter sources need not satisfy $T - 0$. Indeed, the solution for
transverse-traceless perturbations shows no pathology at
$m_i^2=2$.  A pathological behaviour near $m_i^2 =2$ in the spin-2
sector appears because we separated the spin-0 contribution from
the spin-2 contribution. This is due to the degeneracy between
spin-2 and spin-0 perturbations at $m^2=2$, which is explained at
the beginning of \S\ref{Stabilization}. This means that the DGP
model has a natural mechanism to cure the pathology of the massive
gravity theory at $m^2=2$ with $T \neq 0$ by accommodating the
physical spin-0 perturbation at $m^2 =2$ when the mass of spin-2
perturbations crosses $m^2 =2$. However, we have shown that it is
this mechanism that transfers the ghost between spin-2 and spin-0
modes.

To conclude, we confirmed in this paper that it is really
difficult to erase this ghost within the context of standard
linear analysis. However, a different way to avoid appearance of
ghost is proposed in Ref.~\cite{Deffayet}. We think still further
study is necessary to conclude whether this ghost is really
harmful or not.

\acknowledgements
We would like to thank N. Kaloper for valuable comments.
KI thanks Takashi Nakamura for his valuable comments and continuous
encouragement.
We thank Sanjeev Seahra for a careful 
reading of the manuscript.
The discussion during the workshop "Brane-World Gravity: Progress and
Problems"
(Portsmouth, 18 - 29 September 2006) was also useful.
KK is supported by PPARC, and TT is supported by Grant-in-Aid for
Scientific
Research, Nos. 16740141 
and by Monbukagakusho Grant-in-Aid
for Scientific Research(B) No.~17340075. 
This work is also supported in part by the 21st Century COE ``Center for
Diversity
and Universality in Physics'' at Kyoto university, from the Ministry of
Education,
Culture, Sports, Science and Technology of Japan
and also by the Japan-U.K. Research Cooperative Program
both from Japan Society for Promotion of Science.

\appendix

\section{Effective action for spin-2 field}

\subsection{Exact Lagrangian}

To obtain the quadratic action for a massive spin-2 field, the
method with the surest footing involves starting with the
Einstein-Hilbert action with cosmological constant and Fierz-Pauli
mass terms. The Lagrangian up to quadratic order on de Sitter
background is \cite{Higuchi}
\begin{eqnarray}
&&L_h = \alpha \sqrt{-\gamma} \biggl(
    - {1 \over 4} \nabla_\mu h_{\nu\lambda} \nabla^\mu h^{\nu\lambda}
    +{1 \over 2} \nabla_\mu h^{\mu\lambda} \nabla^\nu h_{\nu\lambda}
    +{1 \over 4} \left( \nabla^\mu h - 2\nabla_\nu h^{\mu \nu} \right) \nabla_\mu h \nonumber\\
&&\qquad\qquad\qquad\qquad\qquad\qquad\qquad\qquad\qquad
    -{1 \over 2} \left( h_{\mu \nu} h^{\mu \nu} + {1 \over 2 } h^2\right)
    -{m^2 \over 4} \left( h_{\mu \nu} h^{\mu \nu} -h^2 \right)
                          \biggr) .
\end{eqnarray}
Here the Hubble parameter is set to unity for simplicity.
The overall normalization of the Lagrangian $\alpha$ can be different
from the conventional one. (By redefinition of field,
one can say that the coupling with the matter energy momentum
tensor is unconventional.)
The Lagrangian including coupling to the matter will be given by
\begin{eqnarray}
&&L = L_h + L_m,
\label{APaction}
\end{eqnarray}
with
\begin{eqnarray}
&&L_m = {1 \over 2}\sqrt{-\gamma}\,
 h_{\mu \nu} T^{\mu \nu}.
\end{eqnarray}
Equations of motion derived from the variation of this Lagrangian become
\begin{eqnarray}
&&\alpha \left( X_{\mu \nu} - {m^2 \over 2} \left( h_{\mu \nu}
-\gamma_{\mu \nu} h \right) \right)
+{1\over 2}T_{\mu \nu} =0, \label{EOMX}
\end{eqnarray}
with
\begin{eqnarray}
&&X_{\mu \nu} \equiv {1 \over 2} \left( \Box h_{\mu\nu}
   - \nabla_\mu \nabla^\alpha h_{\alpha\nu}
  - \nabla_\nu \nabla^\alpha h_{\alpha\mu} + \nabla_\mu\nabla_\nu h \right) \nonumber \\
&&\qquad\qquad\qquad\qquad\qquad\qquad
 + {1\over 2} \gamma_{\mu \nu} \left( \nabla_\alpha \nabla_\beta h^{\alpha\beta}- \Box h\right)
 -\left( h_{\mu \nu} + {1\over 2} \gamma_{\mu \nu}h \right).
\end{eqnarray}
Since $\nabla^\mu X_{\mu\nu}=0$ and $\nabla^\mu T_{\mu\nu}=0$, we
have a constraint:
\begin{eqnarray}
\nabla^\mu h_{\mu \nu} = \nabla_\nu h.
\label{APtransverse}
\end{eqnarray}
Substituting back this relation to the trace of Eq.~(\ref{EOMX}),
we obtain
\begin{eqnarray}
h= - {1 \over 3 \alpha (m^2-2)} T.
\label{APtrace}
\end{eqnarray}

Again, substituting the above constraint
equations~(\ref{APtransverse}) and (\ref{APtrace}) into
Eq.~(\ref{EOMX}), we obtain
\begin{eqnarray}
\alpha (\Box^{(4)} -2-m^2 )h_{\mu \nu}& \eq &
  -T_{\mu \nu}
  -{1\over 3(m^2-2)}\left(\nabla_\mu\nabla_\nu
            -(m^2-1)\gamma_{\mu\nu} \right)T\cr
& \eq &   -\left(T_{\mu \nu}-{1\over 4}\gamma_{\mu\nu}T
  +{1\over 3(m^2-2)}\left(\nabla_\mu\nabla_\nu
            -{1\over 4}\gamma_{\mu\nu}\Box^{(4)}\right)T
   \right)
   -{\Box^{(4)}-2-m^2\over 12(m^2-2)}\gamma_{\mu\nu}T.
\label{APhmunu}
\end{eqnarray}
In the last line, we divided 
the traceless part and the pure trace part.
Comparing the solution of the above equations of motion
with the result obtained from the original theory (\ref{soltensor}),
we can read the value of $\alpha$.
The contribution of the last term in (\ref{APhmunu}) is combined
with the second term in Eq.~(\ref{soltensor}) to obtain
\begin{equation}
\frac{2\kappa^2}{3H_+^2}\left( \sum_i \frac{u^2_i(y_+)}{m_i^2 -2}
         \right) \left[ \nabla_\mu \nabla_\nu
        + \gamma_{\mu\nu}
        \right]\frac{1}{\Box^{(4)}+4}T^{(+)}.
\end{equation}
The first term in the square brackets is pure gauge,
while the second term is cancelled
by the pole at $(\Box^{(4)}+4)=0$ in the trace part given in
(\ref{solscalar}) as was shown
in \S\ref{sec:NoCMM}.

\subsection{Simplified Lagrangian}
One may feel like to use the following simplified form of the
effective Lagrangian with the constraints~(\ref{APtransverse})
and (\ref{APtrace}) manifestly imposed by using Lagrange multipliers
$\lambda_\mu$ and $\sigmalambda$:
\begin{eqnarray}
L'_h&\eq &\int d^4x \sqrt{-\gamma}  \Biggl( {\alpha \over 4}
  \sum_i h^{\mu \nu}( \Box^{(4)} -2- m^2 )h_{\mu \nu}
\cr
&& \qquad\qquad\qquad\qquad
 + \left( \nabla_\nu h^{\mu \nu} +\nabla^\mu {1 \over 3 \alpha (m^2-2) )} T \right) \lambda_{\mu}
 + \left( h^{\mu}_{\ \mu} +{1 \over 3 \alpha (m^2-2) )} T\right)
     \sigmalambda \Biggr).
\end{eqnarray}
This Lagrangian is not completely equivalent to $L_h$
as is shown below. Further disadvantage of this Lagrangian
is the appearance of $T^{\mu\nu}$ in the gravitational
part of the Lagrangian.

Equations of motion derived from the variation of $L'_h+L_m$ are
\begin{eqnarray}
&&{\alpha \over 2} (\Box^{(4)} -2-m^2 )h_{\mu \nu}+{1\over 2} T_{\mu
 \nu} -\lambda_{(\mu ;\nu)} + \gamma_{\mu \nu} \sigmalambda=0,
\label{eq9}\\
&&\nabla^\nu h_{\mu \nu} = - \nabla_\mu {1 \over 3 \alpha (m^2-2) )} T ,
\label{eq10}\\
&&h= -{1 \over 3 \alpha (m^2-2) )} T.
\label{eq11}
\end{eqnarray}
Trace of Eq.~(\ref{eq9}) with the aid of Eq.~(\ref{eq11})
gives
\begin{eqnarray}
&&-{\Box^{(4)} -2-m^2\over 6(m^2-2)} T
+{1\over 2} T-\nabla_\mu \lambda^\mu +4\sigmalambda=0.
\label{tra}
\end{eqnarray}
While divergence of Eq.~(\ref{eq9}) becomes
\begin{eqnarray}
&&-\nabla_\mu {\Box^{(4)} +4-m^2 \over 6 (m^2-2)} T
-{1\over 2} \left(\Box^{(4)}\lambda_\mu + \nabla_\nu\nabla_\mu
 \lambda^\nu \right) + \nabla_\mu \sigmalambda=0,
\label{l}
\end{eqnarray}
where we have used the equation
\begin{eqnarray}
\nabla^\mu \Box^{(4)} h_{\mu \nu}
&=& ( \Box^{(4)} + 5 ) \nabla^\mu h_{\mu \nu} -2 \nabla_\nu h \nonumber\\
&=& - (\Box^{(4)} +3) \nabla_\nu {1 \over 3 \alpha (m^2-2)} T \nonumber \\
&=& - \nabla_\nu (\Box^{(4)} +6) {1 \over 3 \alpha (m^2-2)} T.
\end{eqnarray}
Eliminating $\nabla_\mu\sigmalambda$
from the gradient of Eq.~(\ref{tra}) and Eq.~(\ref{l}),
we obtain
\begin{eqnarray}
{1 \over 2} {1 \over (m^2-2)} \nabla_\mu ( \Box^{(4)} +4 ) T
- \nabla_\mu \nabla_\nu \lambda^\nu + 2\Box^{(4)} \lambda_\mu +2
\nabla_\nu \nabla_\mu\lambda^\nu=0.
\end{eqnarray}
Divergence of the above equation becomes
\begin{eqnarray}
{1 \over 2} {1 \over (m^2-2)} \Box^{(4)} ( \Box^{(4)} +4 ) T-
 3(\Box^{(4)} +4) \nabla_\mu \lambda^\mu=0,
\label{key1}
\end{eqnarray}
where we have used
\begin{eqnarray}
\nabla^\mu \Box^{(4)} \lambda_\mu &=& \nabla^\mu\nabla_\nu\nabla_\mu \lambda^\nu \nonumber\\
&=& ( \Box^{(4)} +3) \nabla_\mu \lambda^\mu.
\end{eqnarray}
If we neglect the possibility of adding an arbitrary
solution of the homogeneous equation $(\Box^{(4)}+4)\nabla_\mu\lambda^\mu=0$,
we find
\begin{eqnarray}
\nabla_\mu \lambda^\mu = -{\Box^{(4)} \over 6(m^2-2)} T. \label{nablalamda}
\end{eqnarray}
Substituting this equation into Eq.~(\ref{tra}), we find
\begin{eqnarray}
\sigmalambda= -\frac{m^2 -1}{6( m^2-2)}T.
\label{sigmalambda}
\end{eqnarray}
Substitution of this equation and Eq.~(\ref{nablalamda})
back into Eq.~(\ref{l}) results in
\begin{eqnarray}
(\Box^{(4)} +3) \lambda_\mu = - {(\Box^{(4)} +3) \over 6( m^2-2 )} \nabla_\mu T.
\end{eqnarray}
Again, if we neglect the homogeneous solution, we find
\begin{eqnarray}
\lambda_\mu = - {1 \over 6( m^2-2 )} \nabla_\mu T.
\label{lambdamu}
\end{eqnarray}
If we substitute (\ref{sigmalambda}) and (\ref{lambdamu})
into (\ref{eq9}), we obatain the same equation as Eq.~(\ref{APhmunu}).
However, strictly speaking, this system has propagating modes
corresponding to the pole at $(\Box^{(4)}+4)=0$, which we have neglected
in obtaining Eqs.~(\ref{sigmalambda}) and (\ref{lambdamu}).

As is mentioned earlier, we need to include $T^{\mu\nu}$ in
the gravitational part of Lagrangian $L'_h$.
This is also unsatisfactory. If we remove $T^{\mu\nu}$
from the Lagrangian $L'_h$, we obtain the result with
the pole at $(\Box^{(4)}+4)=0$ even if we neglect
homogeneous solutions.

This discrepancy between the simplified Lagrangian and
the original one is not so strange.
Here what we call constraints are not really constraints.
In the Lagrangian formulation, they are just a part of
equations of motion. In general, it is not always justified
to impose a part of equations of motion before
taking the 
variation of the action.

\subsection{remark on adding matter coupling to the reduced action}

In this paper,
to derive the effective action with source term,
we simplly added the universal coupling term
$(1/2)\sqrt{-\gamma}\, h_{\mu\nu}T^{\mu\nu}$
to the effective Lagrangian valid for the cases without source.
Sometimes justification of this naive appoach is not
so transperent for constrained system.
In the context of cosmological perturbation,
in order to handle constrained system, we often use
the effective action written in terms of the
physical degrees of freedom alone, which is obtained
by imposing all the constraints on the action
to eliminate unphysical degrees of freedom
in the canonical foralism.
In this case, even in the linear theory,
it is not so manifest whether
the resulting equations of motion derived from the
action obtained by adding the coupling term
$(1/2)\sqrt{-\gamma}\, h_{\mu\nu}T^{\mu\nu}$
to the already reduced Lagrangian
are identical to those obtained by taking
the coupling term into account from the beginning.
Here we give a proof of this identity.


First, we briefly describe the method for deriving
the reduced action in the source-free case.
We consider the quadratic Lagrangian $L (q_i ,\dot q_j )$
where $q_i$ represent the dynamical variables.
The corresponding Hamiltonian and the primary constraints are,
respectively, denoted by
$H(q_i ,p_j):= \sum_j p_j\dot q_j-L(q_i, \dot q_j)$
and $C_i =\sum _j \alpha_j q_j + \sum_k \beta_k p_k =0$,
where
$p_j :=\partial L / \partial \dot q_j$.
$\alpha_j$ and $\beta_k$ are, in general, 
given functions determined by the background,
but we here assume that they are just constants for simplicity.
Following the Dirac's prescription \cite{Dirac},
we make a ``new'' Hamiltonian by adding primary
constraints with Lagrange multipliers $\lambda_i$ as
\begin{eqnarray}
H' = H(Q_i ,C_j) + \sum_i \lambda_i C_i,
\label{newH}
\end{eqnarray}
where $Q_i(q_i,p_i)$ are the combinations of variables
which are independent of $C_i$.
From the consistency conditions for the time evolution of constraints,
the secondary constraints follow.
We denote them by $\chi_i$.
Here we assume that the set of constrains $C'_i$ including
both $C_i$ and $\chi_j$ is
second class. Namely the matrix composed of the commutators
between constrains is non-degenerate
\begin{equation}
 \det\left([C'_i,C'_j]\right)\ne 0.
\label{nondegenerate}
\end{equation}
If the set of constraints is not second class,
we add some gauge fixing conditions to make it second class.
Notice that, from the consistency conditions, we also have
equations for $\lambda_i$.
Hence, now $\lambda_i$ are given as functions of
variables $Q'_i$, where $Q'_i$ are the combinations of variables
which are independent of $C'_j$.
(One can add $C'_j$ to $\lambda_i$ without destoying consistency
conditions.)
Thanks to (\ref{nondegenerate}), we can
define $Q'_i$ in such a way that
\begin{equation}
 \left[ Q'_i , C'_j \right] =0.
\label{defQp}
\end{equation}

Here a key point is as follows.
To obtain a consistent set of constraints, in general, we
need to add new term to the Hamiltonian corresponding
to the secondary constraints $\chi_i$ as we did
for the primary constraints in Eq.~(\ref{newH}).
However, in practice, these additional terms are not necessary
in most cases that we are interested in,
although general proof is lacking as far as we know
\footnote{We can prove that these additional terms are not
necessary in linear theory that does not
require additional gauge conditions. This category of
models contains the 
Fierz-Pauli model for massive gravity.}.
Here we assume this property.

We rewrite the Hamiltonian as
\begin{eqnarray}
H' = H (Q'_i , C_j, \chi_k ) + \sum_i \lambda_i(Q'_j) C_i.
\end{eqnarray}
The consistency conditions for the constraints mean that
$
\left[C'_i,H' \right]
$ does not contain $Q'_j$.
This fact with the aid of (\ref{nondegenerate}) and (\ref{defQp})
means that $H'$
does not contain terms in the form $Q'_i C'_j$.
Since $\lambda _i C_i$ has only terms in the form of
$Q'_i C_j$, we find that
$H$ does not have terms in the form of $Q'_i\chi_j$.
The reduced Lagrangian is obtained by setting $C'_i=0$ in
\begin{equation}
 \sum_i p_i(Q'_j,C'_k) \dot q_i(Q'_j,C'_k)-H'(Q'_j,C'_k).
\end{equation}
Here the first kinetic term also does not have any cross-term
between $Q_i$ and $\chi_j$ because $Q_i$ are chosen so that they
satisfy Eq.~(\ref{defQp}).

We are ready to proceed to the setup in which the source
term, $\sum_i q_i S_i$, is present.
To distinguish the quantities defined under the presence
of the source term, we
associate an overbar ``$~\bar{}~$'' with them.
Lagrangian becomes
$\bar L = L(q_i, \dot q_j) +\sum_k q_k S_k$.
Adding source term $\sum_i q_i S_i$ does not affect
the definition of the conjugate momentum and hence
that of the primary constraints. Hence they are identical
to the source-free case,
\begin{equation}
 \bar p_i :=\partial \bar L / \partial \dot q_i =p_i,
\qquad
 \bar C_j =C_j.
\end{equation}
The modification of the Hamiltonian is also trivial.
The ``new'' Hamiltonian with constraint terms becomes
\begin{eqnarray}
\bar H' = H\left( Q'_i , C_j, \chi_k \right) - \sum_\ell q_\ell S_\ell
           + \sum_\ell \bar \lambda_\ell C_\ell .
\end{eqnarray}

As in the source-free case,
secondary constraints ``$\bar \chi_i$'' follow from the consistency
conditions.
Since the limit $S_i\to 0$ should recover the source-free case,
in linear theory the secondary constraints must take the form
\begin{eqnarray}
\bar \chi_i = \chi_i + \sum_j \alpha_{ij} S_j,
\end{eqnarray}
where $\alpha _{ij}$ are some constants.
For the same reason the Lagrange multiplier
``$\bar\lambda_i$''
also become $\bar \lambda_i = \lambda_i + \sum_j \beta_{ij} S_j$.
The reduced Lagrangian is obtained by setting $\bar C'_j=0$ in
\begin{eqnarray}
 &&\sum_n p_n(Q'_i, C_j, \chi_k(\bar\chi_\ell,S_m))
    \dot q_n(Q'_i, C_j, \chi_k(\bar\chi_\ell,S_m))\cr
&&\qquad\qquad\qquad
   - H\left( Q'_i , C_j, \chi_k(\bar\chi_\ell,S_m)
         \right)
    + \sum_n q_n S_n
           + \sum_p \bar \lambda_p(Q'_i,S_j) C_p .
\end{eqnarray}
As we have shown earlier,
the first kinetic term and $H$ have no cross-term
between $Q'_i$ and $\chi_j$.
Therfore, these terms contain no cross-term
between
$Q'_i$ and $S_i$, either.
In the above expression for the Lagrangian,
only the term $\sum_n q_n S_n$ contains the coupling
between $Q'_i$ and $S_i$.
This means that
when we consider the system with the source term,
it is justified to add the source term
to the reduced Lagrangian obtained for the source-free system.

A little subtlety arises
when we reconstruct all the perturbation variables $q_i$
from the physical degrees of freedom $Q'_j$.
To do the reconstruction, we need to solve the constraint equations.
Here we need to recall that we are using a different set of constraints
in two cases with and without source.
As was mentioned above, $\bar \chi_i$ is not identical to
$\chi_i$. In this sense therefore the difference between
these two cases is
not completely captured by the difference in the reduced Lagrangian.
However, what we do in this process is just to solve equations
which does not have temporal differentiation.
Therefore even if we use $\chi_i=0$ in error instead of
$\bar\chi_i=0$ in the case with source, the error is such
that can be absorbed by a shift
of $q_i$ to $q_i+\sum_j \gamma_{ij} S_j$.


\begin{thebibliography}{99}
\bibitem{SN}A. G. Riess et al., Astron. J. \textbf{116}, 1009 (1998).\\
            A. G. Riess et al., Astron. J. \textbf{607}, 665 (2004).
\bibitem{anth}B. J. Carr and M. J. Rees, Nature \textbf{278}, 605 (1979).
\bibitem{const}S. Weinberg, Rev. Mod. Phys. \textbf{61}, 1 (1989).
\bibitem{quint}C. Wetterich, Nucl. Phys. \textbf{B302}, 668 (1988).\\
               B. Ratra and P. J. E. Peebles, Phys. Rev. \textbf{D37}, 3406 (1988).
\bibitem{FP}M. Fierz and W. Pauli, Proc. Roy. Soc. \textbf{173}, 211 (1939).
\bibitem{Higuchi}A. Higuchi, Nucl. Phys. \textbf{B282}, 397 (1987).
\bibitem{DGP}G. R. Dvali, G. Gabadadze and M. Porrati, Phys. Lett. \textbf{B485}, 208 (2000).
\bibitem{Lue}A. Lue, Phys. Rept. \textbf{423}, 1-48 (2006).
\bibitem{self}C. Deffayet, Phys. Lett. \textbf{B502}, 199 (2001).
\bibitem{SSS}T. Tanaka, Phys. Rev. \textbf{D69}, 024001 (2004).\\
             C. Middleton and G. Siopsis, Mod. Phys. Lett. \textbf{A19}, 2259 (2004).\\
             G. Gabadadze and A. Iglesias, Phys. Rev. \textbf{D72} 084024 (2005).
\bibitem{Lue2}
A.~Lue and G.~Starkman, Phys.\ Rev.\ D {\bf 67} (2003) 064002
A.~Lue, R.~Scoccimarro and G.~D.~Starkman, Phys.\ Rev.\ D {\bf 69} (2004) 124015.
\bibitem{SF}K. Koyama and R. Maartens, JCAP \textbf{0601}, 016 (2006).\\
I.~Sawicki, Y.~S.~Song and W.~Hu, arXiv:astro-ph/0606285.\\
Y.~S.~Song, I.~Sawicki and W.~Hu, arXiv:astro-ph/0606286.
\bibitem{strong}N. Arkani-Hamed, H. Georgi and M. D. Schwartz, Annals Phys. \textbf{305}, 96 (2003).\\
                V. A. Rubakov, arXiv:hep-th/0303125 (2003).
\bibitem{SW}N. Kaloper, Phys. Rev. Lett. \textbf{94}, 181601 (2005).\\
            N. Kaloper, Phys. Rev. \textbf{D71}, 086003 (2005).
\bibitem{SG}M. A. Luty, M. Porrati and R. Rattazzi, JHEP \textbf{0309}, 029 (2003).\\
            A. Nicoils and R. Rattazzi, JHEP \textbf{0406}, 059 (2004).
\bibitem{Koyama}K. Koyama, Phys. Rev. \textbf{D72}, 123511 (2005). \\
                C. Charmousis, R. Gregory, N. Kaloper and A. Padilla, arXiv:hep-th/0604086 (2006).
\bibitem{Koyama2} D. Gorbunov, K. Koyama and S. Sibiryakov, Phys. Rev. \textbf{D73}, 044016 (2005).
\bibitem{GW}W. D. Goldberger and M. B. Wise, Phys. Rev. Lett. \textbf{83} 4922 (1999).\\
            W. D. Goldberger and M. B. Wise, Phys. Lett. \textbf{B475} 275 (2000).
\bibitem{TM}T. Tanaka and X. Montes, Nucl. Phys. \textbf{B528}, 259 (2000).
\bibitem{Deser}
S.~Deser and R.~I.~Nepomechie, Ann. Phys. {\bf 154} (1984) 396. \\
S.~Deser and A~Waldron, Phys. Lett. {\bf B508} (2001) 347.
\bibitem{TT}J. Garriga and T. Tanaka, Phys. Rev. Lett. \textbf{84}, (2000).
\bibitem{Kaloper}In JGRG-15 meeting, K.I and T.T made an incorrect claim that the ghost should disappear in
two-branes model. It was pointed out by N.~Kaloper during the meeting that the ghost should remain in this
setup.
\bibitem{open} {\it See, e.g., appendix B in}
  J.~Garriga, X.~Montes, M.~Sasaki and T.~Tanaka,
  Nucl.\ Phys.\ B {\bf 513}, 343 (1998).
\bibitem{Deffayet}
  C.~Deffayet, G.~Gabadadze and A.~Iglesias,
  JCAP {\bf 0608}, 012 (2006).


\end{thebibliography}
\end{document}